\documentclass[aps,prl,twocolumn,superscriptaddress,showpacs]{revtex4-2}
\def\cm{cm$^{-1}$}
\usepackage{graphicx}
\usepackage{sidecap}
\sidecaptionvpos{figure}{t}
\usepackage{color}
\usepackage{epstopdf}
\usepackage{amssymb}
\usepackage{amsmath}
\usepackage{amsfonts}
\definecolor{darkblue}{rgb}{0,0.02,0.45}
\def\cdbl{\color{darkblue}}
\usepackage{color}
\usepackage[colorlinks,bookmarks=false,citecolor=darkblue,linkcolor=red,urlcolor=blue]{hyperref} 
\definecolor{darkred}{rgb}{0.7,0.0,0.0}

\definecolor{darkblue}{rgb}{0,0.02,0.45}
\def\cdbl{\color{darkblue}}
\definecolor{darkgreen}{rgb}{0.02,0.45,0.0}

\definecolor{violet}{rgb}{0.8,0.2,0.6}

\begin{document}
\title{Phonon and magnon dynamics across antiferromagnetic transition\\ in 2D layered van der Waals material CrSBr}

\author{E. Uykur}
\email{e.uykur@hzdr.de}
\affiliation{Helmholtz-Zentrum Dresden-Rossendorf, Ion Beam Physics and Materials Research, 01328 Dresden, Germany}

\author{A. A. Tsirlin}
\affiliation{Felix Bloch Institute for Solid-State Physics, Leipzig University, 04103, Leipzig, Germany}

\author{F. Long}
\affiliation{Helmholtz-Zentrum Dresden-Rossendorf, Ion Beam Physics and Materials Research, 01328 Dresden, Germany}
\affiliation{Institute of Applied Physics, TUD Dresden University of Technology, 01062 Dresden, Germany}

\author{M. Wenzel}
\affiliation{1.~Physikalisches Institut, Universität Stuttgart, D-70569, Stuttgart, Germany}

\author{M. Dressel}
\affiliation{1.~Physikalisches Institut, Universität Stuttgart, D-70569, Stuttgart, Germany}

\author{K. Mosina}
\affiliation{Department of Inorganic Chemistry, University of Chemistry and Technology Prague, Technická 5, 166 28 Prague 6, Czech Republic}

\author{Z. Sofer}
\affiliation{Department of Inorganic Chemistry, University of Chemistry and Technology Prague, Technická 5, 166 28 Prague 6, Czech Republic}

\author{M. Helm}
\affiliation{Helmholtz-Zentrum Dresden-Rossendorf, Ion Beam Physics and Materials Research, 01328 Dresden, Germany}
\affiliation{Institute of Applied Physics, TUD Dresden University of Technology, 01062 Dresden, Germany}

\author{S. Zhou}
\affiliation{Helmholtz-Zentrum Dresden-Rossendorf, Ion Beam Physics and Materials Research, 01328 Dresden, Germany}

\date{\today}

\begin{abstract}
We report temperature-dependent reflectivity spectra of the layered van der Waals magnet CrSBr in the far-infrared region. Polarization-dependent measurements resolve the vibrational modes along the E$\|a$- and $b$-axes and reveal the clear structural anisotropy. While the $a$-axis phonons notably harden on cooling, the $b$-axis phonon frequencies are almost temperature-independent. A phonon splitting due to the antiferromagnetic phase transition is observed for the 180~\cm\ $a$-axis vibrational mode, accompanied by a phonon softening below $T_N$. Furthermore, an additional mode with strong magnetic characteristics at $\sim$360~\cm\ is identified and attributed to the magnon excitation of CrSBr.   
\end{abstract}

\maketitle

{\cdbl\textit{Introduction.}} 

As an ideal playground for low-dimensional magnetism, two-dimensional (2D) van der Waals (vdW) magnets attract significant attention. The long-range magnetic order is facilitated by the magnetic anisotropy and/or dipolar interactions in these materials. The possibility of exfoliation and integration into heterostructures and multifunctional devices makes them very attractive for potential applications, such as spintronics, magneto-optical devices, and quantum computing, etc. After the first reports of 2D ferromagnetism in CrI$_3$~\cite{Huang2017} and Cr$_2$Ge$_2$Te$_6$~\cite{Gong2017}, significant efforts have been devoted to the discovery of the many others. The MPX$_3$ family~\cite{Jiang2021} was proposed as an air-stable alternative, but these compounds are typically antiferromagnetic. 

Very recently, CrSBr regained interest due to its interesting properties~\cite{Goser1990}: CrSBr is an antiferromagnetic semiconductor with the bulk N\'eel temperature of around $T_N = 132$~K. Monolayer CrSBr is ferromagnetic, and its ordering temperature might be higher than 132~K~\cite{Guo2018,Wang2020,Yang2021}. It is air stable even in the monolayer form, which makes this compound specifically attractive for potential applications. 

Under ambient conditions, CrSBr possesses an ortho-rhombic $Pmmn$ structure. The crystal structure is formed by two Cr-layers bounded by the S-atoms. Each layer is terminated by the Br-atoms, which are coupled to the next layers with weak van der Waals bonds. The magnetic structure of CrSBr is typically referred to as $A$-type where each Cr layer is ferromagnetic, whereas the antiferromagnetic order is established between the adjacent layers. Earlier studies reported an additional ferromagnetic phase transition at around 40~K accompanied by changes in magnetoresistance~\cite{Telford2020,Telford2022,LopezPaz2022,Wu2022}. However, later studies~\cite{Long2023} on high-quality single crystals demonstrated the absence of this low-temperature phase suggesting the extrinsic origin, such as disorder. This assumption is also corroborated by the recent studies on irradiated samples~\cite{Long2023b}.

CrSBr is a narrow-band semiconductor with a direct band gap of around 1.3~eV (in bulk)~\cite{Telford2020,Yang2021,Wilson2021} making it suitable for the opto-electronic applications. The band structure and the magnetic properties are shown to be highly anisotropic~\cite{Wu2022, Cham2022,Klein2023}. Manipulation of magnetic materials with light is among the current trends in spintronics. The relatively high energies of the magnon excitations in the MPX$_3$ family led to unusual effects associated with magnon-phonon coupling and magnon polarons~\cite{Vaclavkova2021,Jana2023,Jana2023b}. In conjunction, the unusual phonon activity is reported for the CrSBr, where the photoluminescence measurements revealed the exciton-phonon coupling in the compound~\cite{Lin2024}. This coupling is shown to be sensitive to the antiferromagnetic transition suggesting that the optical response of CrSBr is strongly dependent on to the underlyig magnetic structure. In fact, the control of the excitons and exciton-magnon coupling is demonstrated by means of cavity photons~\cite{Dirnberger2023} and ultrafast non-equilibrium dynamics~\cite{Bae2022,Meineke2024}, as well.

\begin{figure*}
\centering
\includegraphics[width=1\linewidth]{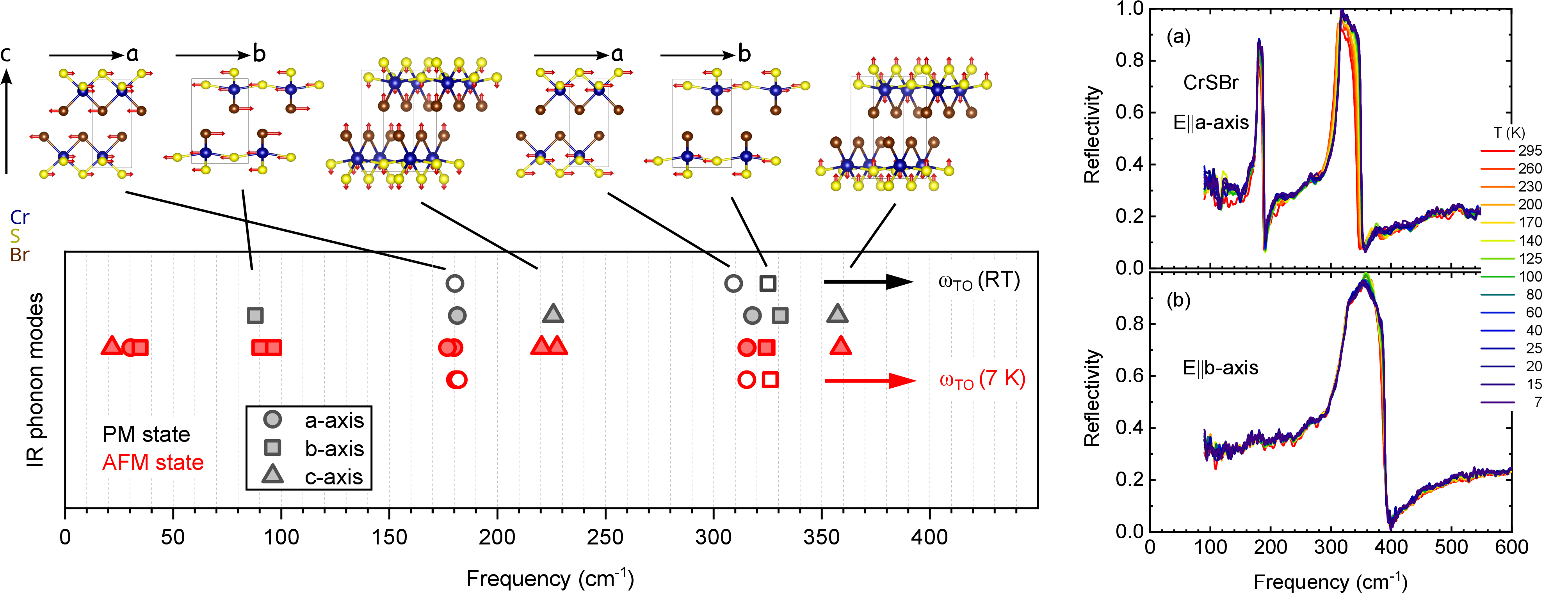}%
\caption{Left panel: Calculated infrared active phonon modes in the paramagnetic (gray symbols) and antiferromagnetic (red symbols) states. Experimental values for the observed modes are also shown with the open symbols at room temperature (reflecting the paramagnetic state) and at 7~K (reflecting the antiferromagnetic state). Points are symbol-coded for the different crystallographic directions. Temperature-dependent reflectivities along (a) $a$-axis and (b) $b$-axis. A clear structural anisotropy is visible. Our experiments probe the in-plane response; therefore, the modes along the $c$-axis are not accessible.  }
\label{F1}%
\end{figure*}

Despite several studies revealing the exotic properties of this 2D vdW antiferromagnet, the (magneto)optical response of CrSBr is mostly investigated by means of photoluminescence spectroscopy~\cite{Moros2023,Pawbake2023,Lin2024} in the visible energy range or with Raman spectroscopy in the frequency range of lattice vibrations~\cite{Lee2021,Klein2022,Torres2023,Linhart2023}. Surprisingly, the infrared-active modes have not been studied, so far. We fill this gap by performing the temperature-dependent infrared spectroscopy study in the far-infrared region. Our results clearly demonstrate the infrared-active modes and the sensitivity of these modes to the antiferromagnetic ordering. We further reveal the structural anisotropy along with the unusual phonon activities. Furthermore, a magnetic mode is found at around 360~\cm, which is attributed to the magnon excitation of the compound.

{\cdbl\textit{Methods.}}

Single crystals of CrSBr are grown with the method explained elsewhere~\cite{Long2023b}. Temperature-dependent reflectivity measurements are performed on the bulk single crystal between 7-295~K in the frequency range of 90-600 \cm. Sample thickness was $\sim$120~$\mu$m and the measurement area was around 80$\times$240~$\mu$m. 

A Hyperion IR microscope coupled to a Bruker Vertex80v is used for the measurements. The sample is placed into a Cryovac microstat and a polarized light is used to probe the E$\|a$- and $b$-axis response of the sample. We used 0.5~\cm\ resolution for the study. 

$\Gamma$-point phonon frequencies are calculated in VASP~\cite{vasp1,vasp2} using the Perdew-Burke-Ernzerhof exchange-correlation potential for solids~\cite{pbesol} and the DFT+$U$ correction for correlation effects in the Cr $3d$ shell with the on-site Coulomb repulsion $U_d=3$ eV, Hund's coupling $J_d=1$ eV, and double-counting correction in the atomic limit~\cite{janson2013,janson2014}. The previously observed A-type antiferromagnetic order was adopted with the unit cell doubled along the $c$ direction. Experimental atomic positions at 1.8 K from Ref.~\cite{LopezPaz2022} are chosen as the starting point and fully optimized prior to phonon calculations. The $12\times 12\times 3$ $k$-mesh is used. The calculated IR-active modes are given in Fig.~\ref{F1}. Our calculations revealed 6 IR active modes in the paramagnetic state. For the antiferromagnetic calculations doubling of the unit cell creates nonequivalent atomic positions in the adjacent layers causing the splitting of the modes. While the splitting of the high energy modes is negligible, a sizable splitting is observed for the low energy modes and also resolved in our experiments.

{\cdbl\textit{Results and Discussion.}}

\begin{figure*}[ht]
\centering
\includegraphics[width=1\linewidth]{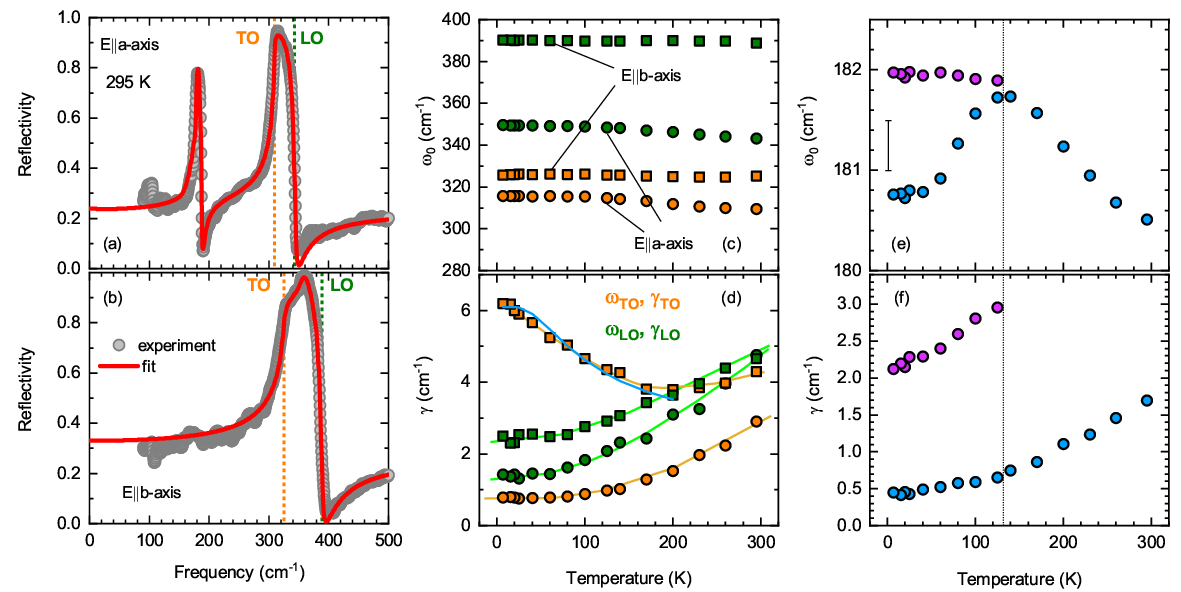}%
\caption{Example for Lorentz fits of the spectra at room temperature (a,b) and the obtained resonance frequencies (c) and scattering rates (d) for the higher-energy transverse and longitudinal modes. Orange symbols are for the transverse modes and the green symbols represent the longitudinal modes. Squares are for the $b$-axis modes and the circles are for the $a$-axis modes. The $a$-axis modes show a larger phonon hardening compared to the $b$-axis modes. Furthermore, although the scattering of the modes exhibits the expected decrease with decreasing temperature, the $b$-axis transverse mode reveals an unusual increasing behavior at lower temperatures indicating an unusual behavior of the mode. (e) and (f) display the parameters of the low-energy $a$-axis mode. This mode shows a clear splitting below the antiferromagnetic transition temperature as labeled with the dashed lines. The scale in (e) shows the experimental resolution. Even though the splitting is small, it is still larger than our resolution.  }
\label{F2}%
\end{figure*}

Figure ~\ref{F1}(a) and (b) shows the temperature-dependent reflectivity along E$\|a$- and $b$-axes, respectively. The far-infrared range depicts the observed infrared-active phonon modes clearly. Our phonon calculations reveal six IR-active phonon modes in total (as demonstrated in Fig.~\ref{F1}), where two of these are $c$-axis modes and hence cannot be observed in our experiments. The low-energy $b$-axis mode is also not visible possibly due to the lower energy limit of our measurement. Relatively good agreement is obtained between the experimental and calculated modes for the other vibrations. Details reveal that the high-energy modes appear as a Reststrahlen band with 100\% reflectivity in our experiments. 

To discuss the observed modes, we modeled the reflectivity with Lorentzians by using the dielectric function given in Eq.(~\ref{Lorentzian}) and extracted the resonance frequencies and the scattering rates of each individual phonon. 

\begin{equation}
\varepsilon(\omega)=\varepsilon_{\infty}\times\prod_i \frac{\omega_{LO,i}^2-\omega^2-i\omega\gamma_{LO,i}}{\omega_{TO,i}^2-\omega^2-i\omega\gamma_{TO,i}}
\label{Lorentzian}
\end{equation}

Here, $\varepsilon_\infty$ is the high-frequency dielectric constant, $\omega_{TO}$ and $\omega_{LO}$ are the transverse and longitudinal frequencies, and $\gamma_{TO}$ and $\gamma_{LO}$ are the scattering rates. Transverse (TO) and longitudinal(LO) phonon frequencies are identified as the left and the right edge of the Reststrahlen band, respectively [Fig.~\ref{F2}(a) and (b)]. The TO frequency corresponds to the resonance with light, because the electromagnetic waves are transverse, and should be compared with DFT.  The electromagnetic waves with frequencies between $\omega_{TO}$ and $\omega_{LO}$ will not propagate at all in the medium but will instead have 100\% reflectivity, in case there is no scattering. This also suggests that significantly thinner samples would be needed to reveal the phonon modes in transmission in this frequency range as by definition the reflectivity between transverse and the longitudinal optical modes will be 100 percent. 

Figure~\ref{F2}(a) and (b) show the spectra at room temperature along with the modeled reflectivity curves for $a$- and $b$-axes. In Fig.~\ref{F2}(c), the resonance frequencies for TO and LO modes are given for both axes. The hardening of the phonon modes with decreasing temperature is observed for all modes, albeit it is much larger for the $a$-axis case. This indicates a much higher sensitivity of the $a$-axis mode to the unit-cell volume and, therefore, also to pressure. In Fig.\ref{F2}(d), the scattering rate of the phonon modes is given. Almost all the modes show, as expected, a gradual decrease of the scattering rate on cooling; whereas the TO mode along $b$-axis reveals an unusual broadening below $\sim$150~K. Such broadening is usually interpreted as the coupling of phonon to the electronic background and in this case, seems to be mediated with the magnetic ordering of the compound. 

By using the electron-phonon coupling scenario~\cite{Bonini2007}, we model the temperature-dependent linewidth of the TO mode by using the Eq.(~\ref{ep}) as shown with the blue curve in Fig.~\ref{F2}(d):

\begin{equation}
\gamma^{e-ph}(T)=\gamma^{e-ph}(0)\left[f\left(-\frac{\hbar\omega_0}{2k_BT}\right)-f\left(\frac{\hbar\omega_0}{2k_BT}\right)\right]
\label{ep}
\end{equation} 

\noindent Here, $\hbar\omega_0 = 325$~\cm\ that is the TO phonon energy, $k_B$ is the Boltzmann constant, and $f(x) = 1/[e^x + 1]$. The intrinsic linewidth, $\gamma^{e-ph}(0)$ = 6.3\cm. 

In Fig.~\ref{F2} (e) and (f), the phonon parameters for the low-energy $a$-axis mode are given. This mode shows a small but clear splitting below the magnetic transition along with a small softening. This splitting is also predicted by our phonon calculations. The linewidth of these modes does not show any anomalies and reveals the expected decrease upon cooling.

\begin{figure*}
\centering
\includegraphics[width=1\linewidth]{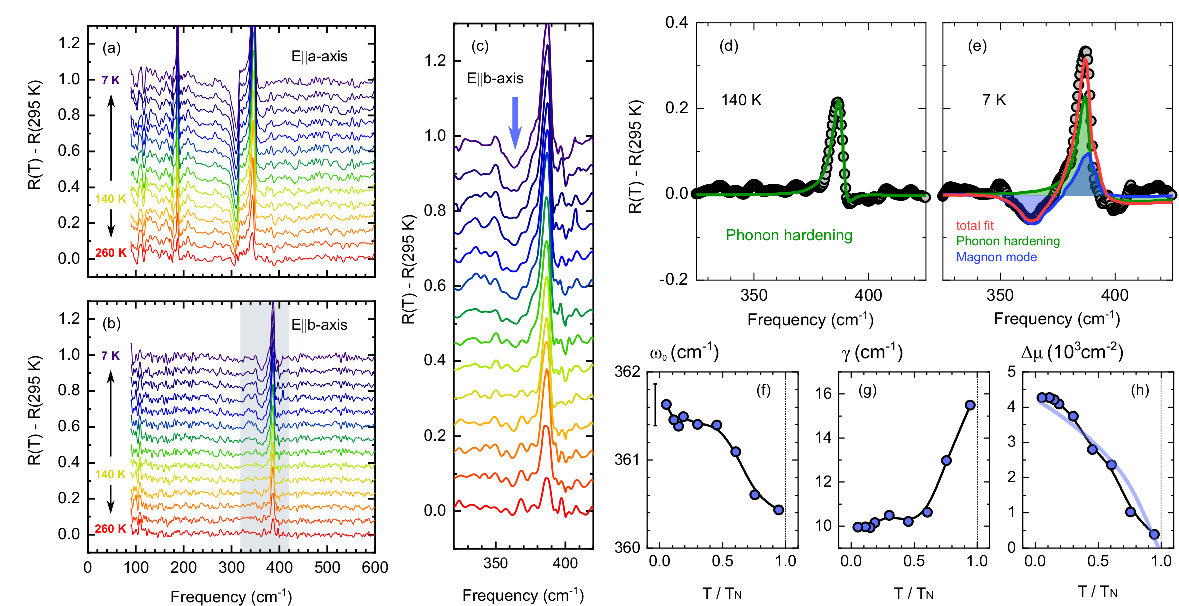}%
\caption{(a) and (b) difference reflectivity spectra as a function of temperature, where the base temperature is chosen as the room temperature spectra for $a$- and $b$-axes. The curves are displaced for clarity reasons. Gray shaded area highlights the mode appearing below $T_N$, which is enlarged in (c). The arrow in (c) points to the magnon mode. (d) and (e) present the fit of the $b$-axis spectra above and below the antiferromagnetic transition, respectively. The contributions for the phonon hardening and the magnon mode are depicted with the shaded areas. (f-h) Fit parameters for the magnetic mode. Scale in (f) is the measurement resolution. The solid line in (h) is the order-parameter-like curve for the eye.     }
\label{F3}%
\end{figure*}

To identify the subtle changes in spectra, we extract the difference reflectivity in Fig.~\ref{F3} (a) and (b) for both $a$- and $b$-axes, respectively. With this method, we can identify the changes related to the phonon modes (phonon hardening) and any additional features appearing below the magnetic transition more easily. As can be seen from the spectra, the major changes occur due to the phonon hardening with decreasing temperature, as expected. However, our $b$-axis spectra also reveal an additional feature below 140~K that cannot be solely attributed to the phonon-mediated changes. In Fig.~\ref{F3}(c), we highlight this additional mode with an arrow. We should also emphasize that any changes related to the TO-vibration fall out of the given energy range and therefore, this additional feature cannot be explained as another phonon hardening as would be seen in $a$-axis spectra. We further examine the mode in Fig.~\ref{F3}(d) and (e), where the difference above and below the transition temperature is demonstrated. We choose 140~K and 7~K spectra as references for the temperatures right above the magnetic transition and deep into the ordered state, respectively. For 140~K, a single peak structure is visible (with a small dip on the higher-energy side) that can be solely accounted for the phonon hardening. On the other hand, for the 7~K spectrum, the situation is more complicated. The phonon hardening is still visible. However, now a dip-peak structure is seen, suggesting a second feature present in the antiferromagnetic state. At first glance, the dip-peak structure is rather unusual, nonetheless can be understood by the magnetic nature of this mode. Unlike dielectric features, the magnetic features in the reflectivity spectra are expected to show the dip-peak structure~\cite{Haussler1982}. Considering the overlapping magnetic mode and the phonon hardening, we can reproduce the difference spectra satisfactorily as shown with the red curve in Fig.~\ref{F3}(e). Please note that the unambiguous determination of the small changes with temperature is not very easy in this case; however, we still can extract the general temperature-dependent parameters of this magnetic mode. 

In Fig.~\ref{F3}(f-h) we plot the extracted parameters of the magnetic mode. The resonance frequency is determined to be $\sim$360~\cm ($\sim$44.6 meV) with a slight hardening upon decreasing temperature. This energy range is very close to the reported optical magnon mode for CrSBr determined by complementary magneto-transmission~\cite{Pawbake2023} and inelastic neutron scattering~\cite{Scheie2022} experiments. As the calculations~\cite{Esteras2022,Bo2023} for monolayer and bulk CrSBr show different values for the magnon modes, we believe that the small discrepancies between certain reports can be attributed to the different sample thicknesses. The width of the magnon mode increases when getting closer to the magnetic transition [Fig.~\ref{F3}(g)] suggesting the presence of strong magnetic fluctuations before the long-range order. Furthermore, we noticed that it is still significantly broad even deep in the ordered state. This also goes hand in hand with the broadening of the $b$-axis optical mode and perhaps is related with the interplay of these two. In Fig.~\ref{F3}(h), we plotted the $\Delta\mu$ parameter which is the intensity of the magnetic mode and represents the change of the magnon population with temperature. It shows an order parameter-like change. 

Our data indicate a strong in-plane anisotropy of CrSBr. While the $a$-axis mode shows a strong hardening upon cooling, the $b$-axis mode remains almost temperature-independent, yet it couples to the magnon. This difference can be ascribed to the peculiar structure of the CrSBr layer that features edge-sharing coupling along `$a$' and corner-sharing coupling along `$b$' between the Cr polyhedra. One interesting question for further investigation is the exact microscopic process that couples the optical magnon to IR light. This coupling most likely involves the phonon.

{\cdbl\textit{Conclusions.}}

In summary, we present here the temperature-dependent phonon and magnon dynamics of CrSBr revealed via infared spectroscopy study. Three in-plane phonon modes are identified. The gradual hardening of the phonon modes is shown in both orientations albeit it is much stronger for the $a$-axis modes. Decreasing of the scattering with lowering temperature is observed for all the modes except the $b$-axis transverse mode suggesting the interplay of this mode with other degrees of freedom. We identified the subtle effects at the magnetic transition, as well. The low-energy phonon split and show a soft mode below T$_N$. Furthermore, a mode with magnetic characteristics appears in the $b$-axis spectra and is located at around 360~\cm. We attributed this feature to an optical magnon. The temperature dependence of this mode reveals the change of the magnon population with temperature. We also noticed that the mode is relatively broad and remains broad even in the ordered state pointing a possible interplay between this mode and phonons.

\begin{acknowledgments}
The authors acknowledge the fruitful discussions with Kaiman Lin and the technical support by Gabriele Untereiner. Z.S. was supported by ERC-CZ program (project LL2101) from Ministry of Education Youth and Sports (MEYS) and by the project Advanced Functional Nanorobots (reg. No. CZ.02.1.01$/$0.0$/$0.0$/$15\_003$/$0000444 financed by the EFRR). M.W. is supported by IQST Stuttgart/Ulm via a project funded by the Carl Zeiss Stiftung. Computations for this work were done (in part) using the resources of the Leipzig University Computing Center.
\end{acknowledgments}

\bibliography{CSBreferences.bib}

\end{document}